\documentclass[useAMS,usenatbib]{mn2e}
\usepackage{graphicx}
\usepackage{amsmath}
\usepackage{dsfont}

\usepackage{amssymb}
\usepackage{amsmath} %, amsthm, amssymb}
\usepackage{subfigure} % \geometry{landscape}
\usepackage{epsfig}
\usepackage{appendix}
\usepackage{xspace}
\usepackage{multirow}
\usepackage{mathptmx}
\usepackage{url}
\usepackage{xcolor}
\usepackage{footnote}
\usepackage{natbib}
\usepackage[colorlinks=false,linkcolor=black, citecolor=green, linktoc=page, urlcolor=cyan, 
pdffitwindow=false]{hyperref}
\usepackage{graphicx}
\usepackage{wrapfig}
\usepackage{epstopdf}
\usepackage{acronym}
\def\reff@jnl#1{{\rm#1\/}}
\def\aj{\reff@jnl{AJ}}         % Astronomical Journal
\def\araa{\reff@jnl{ARA\&A}}      % Annual Review of Astron and Astrophys
\def\apj{\reff@jnl{ApJ}}        % Astrophysical Journal
\def\apjl{\reff@jnl{ApJ}}        % Astrophysical Journal, Letters
\def\apjs{\reff@jnl{ApJS}}       % Astrophysical Journal, Supplement
\def\aap{\reff@jnl{A\&A}}        % Astronomy and Astrophysics
\def\aapr{\reff@jnl{A\&A~Rev.}}     % Astronomy and Astrophysics Reviews
\def\aaps{\reff@jnl{A\&AS}}       % Astronomy and Astrophysics, Supplement
\def\mnras{\reff@jnl{MNRAS}}      % Monthly Notices of the RAS
\def\physrep{\reff@jnl{Physics Reports}}% Physics Reports
\def\prd{\reff@jnl{Phys.Rev.D}}     % Physical Review D
\def\prl{\reff@jnl{Phys.Rev.Lett}}   % Physical Review Letters
\def\pasp{\reff@jnl{PASP}}       % Publications of the ASP
\def\pasj{\reff@jnl{PASJ}}       % Publications of the ASJ
\def\nat{\reff@jnl{Nature}}       % Nature
\def\jcap{\reff@jnl{JCAP}}   %Journal of Cosmology and Astroparticle Physics
\def\memsai{\reff@jnl{MemSAI}} %Memorie della Societa Astronomica Italiana Supplement 
\def\na{\reff@jnl{New Astronomy}}       % New Astronomy

\newcommand{\simgt}{\lower.5ex\hbox{$\; \buildrel > \over \sim \;$}}
\newcommand{\simlt}{\lower.5ex\hbox{$\; \buildrel < \over \sim \;$}}
\newcommand{\bm}[1]{\mbox{{\it \boldmath$#1$}}}

\usepackage{color}

\def\Sref#1{$\S$\ref{#1}\xspace}

\def\Fref#1{Figure~\ref{#1}\xspace}

\def\Eref#1{Equation~\ref{#1}\xspace}

\def\Aref#1{Appendix~\ref{#1}\xspace}
\def\Cref#1{Chapter~\ref{#1}\xspace}

\def\upenn{Department of Physics and Astronomy, Center of Particle Cosmology, 
University of Pennsylvania, 209 South 33rd Street, Philadelphia, PA 19104, USA}
\def\ethz{ETH Zurich, Department of Physics, Wolfgang-Pauli-Strasse 27, 
8093 Zurich, Switzerland}

\begin{document}
\twocolumn 
\title{Delensing Galaxy Surveys}
\author[C. Chang and B. Jain]{
Chihway Chang,$^{1}$ 
Bhuvnesh Jain$^{2}$  \\
$^{1}$\ethz\\
$^{2}$\upenn\\ } 
 
 \date\today
\maketitle

\begin{abstract}

Weak gravitational lensing can cause displacements, magnification, rotation and shearing of the images of 
distant galaxies. Most studies focus on the shear and magnification effects since they are more 
easily observed. In this paper we focus on the effect of lensing displacements on wide field images. Galaxies 
at redshifts 0.5--1 are typically displaced by 1 arcminute, and the displacements are coherent over 
degree-size patches. However the displacement effect is redshift-dependent, so there is a visible relative 
shift between galaxies at different redshifts, even if they are close on the sky. We show that the 
reconstruction of the original galaxy position is now feasible with lensing surveys that cover many hundreds 
of square degrees. We test with simulations two approaches to ``delensing'': one uses shear measurements 
and the other uses the foreground galaxy distribution as a proxy for the mass. We also estimate the effect of foreground deflections on galaxy-galaxy lensing measurements and find it is relevant only for LSST and Euclid-era surveys. 

\end{abstract}


\section{Introduction}

The deflection of light rays by intervening structures, a phenomenon referred to as gravitational lensing, 
provides astronomers with a unique tool to study the mass distribution in the universe. Unlike 
other observational probes, lensing  provides a direct measure of the mass, including dark matter, 
irrespective of the dynamical state \citep{1998ApJ...498...26K, 2001PhR...340..291B, 2003ARA&A..41..645R, 
2008ARNPS..58...99H} of the mass distribution. The first-order effect of weak lensing on distant galaxies is 
a displacement of the observed position compared to its true position in the sky. The second-order effect, 
due to the difference in the deflection between light rays from the same galaxy, causes a shear and 
magnification of the observed galaxy. 

The first-order displacement effect, albeit physically large compared to shear and magnification, is hard to 
observe since the true positions of the sources are unknown. On the other hand, the second-order effects, 
shear and magnification, have been measured with reasonable accuracy in existing data on distant 
galaxies \citep{2011arXiv1111.1070H, 2012MNRAS.427..146H, 2013ApJ...765...74J}. 
However for the cosmic microwave background (CMB), it is the lensing displacement that is used to 
reconstruct the mass distribution \citep{2006PhR...429....1L}. The CMB temperature field is a smooth 
field well described as a Gaussian random field, which enables the mass reconstruction.  
\citet{2007PhRvD..75j3509V} and \citet{2008PhRvD..78d3508D} have investigated the impact of weak lensing 
displacements on the galaxy power spectrum, in particular on measurements of the baryon acoustic 
oscillation (BAO) peaks, and found that the effect is generally too small to impact existing data.    

The main goal of this paper is to revisit the the impact of weak lensing deflections on the spatial distribution 
of galaxies, and describe a practical framework to reconstruct the displacement field using observational 
quantities in galaxy surveys. This work is especially relevant for ongoing and future wide-field surveys 
such as the Kilo-Degree Survey\footnote{\url{http://kids.strw.leidenuniv.nl/}}, 
the Hyper SuprimeCam survey\footnote{\url{http://www.naoj.org/Projects/HSC/}}, the Dark Energy 
Survey\footnote{\url{http://www.darkenergysurvey.org/}}, the LSST survey\footnote{\url{http://www.lsst.org/lsst/}} 
and the Euclid survey\footnote{\url{http://sci.esa.int/euclid/}}, 
where the large sky coverage allows more accurate reconstruction of the deflection field. 

The paper is organised as followed. In \Sref{sec:theory} we describe the basic formalism associated with the 
weak lensing deflection field calculation. In \Sref{sec:observation} we discuss the observational consequences 
of the weak lensing deflection. As an example we calculate the error introduced by the deflection on galaxy-galaxy 
lensing measurements in \Sref{sec:g-g}. The reconstruction framework is laid out and tested in 
\Sref{sec:reconstruction}. We conclude in \Sref{sec:conclusion}.  

\section{Formalism}
\label{sec:theory}

 \begin{figure*}
\begin{center}
  \includegraphics[scale=0.45]{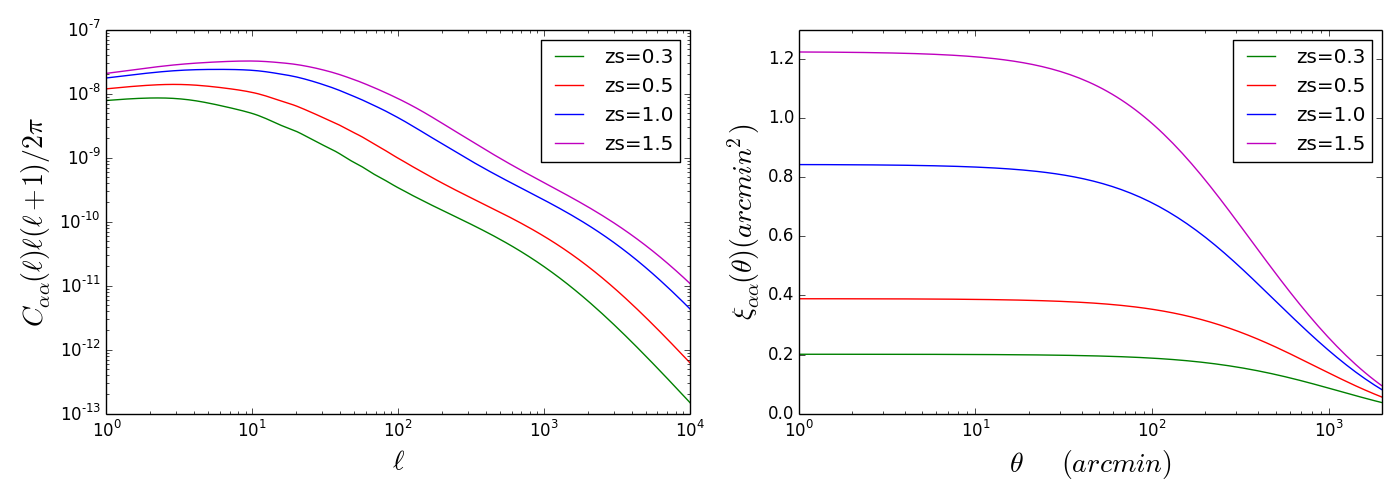}
  \caption{The power spectrum (left) and correlation function (right) for the deflection angle $\alpha$ due to 
  weak lensing of sources at four different redshifts. The power  peaks at $\ell<100$ for 
  these redshifts, which corresponds to an angular scale above 1 degree. The correlation function is shown in 
  units of arcmitute$^{2}$.}
  \label{fig:Cl_alpha}
\end{center}
\end{figure*}

In the weak lensing regime, galaxy images are distorted. The distortion is effectively a mapping from the source 
plane to the image plane, which can be described by the lensing Jacobian $A$:
\begin{equation}
A = \begin{pmatrix}
 1-\kappa - \gamma_1 & - \gamma_2 \\
-\gamma_2 & 1-\kappa+\gamma_1
\end{pmatrix}, 
\label{eq:jacobian}
\end{equation}
where $\kappa$ is the convergence and $\gamma_i$ is the two-component shear. The convergence 
leads to a magnification of the object's size and the shear causes an initially circular object to appear 
elliptical. The convergence $\kappa$ is a scalar quantity and is given by a weighted projection of the 
mass density fluctuation field:
\begin{equation}
\kappa(\bm{\theta})=\frac{1}{2}\nabla^2\Psi(\bm{\theta})
=\int\!\!d\chi W(\chi) 
\delta[\chi, \chi\bm{\theta}].
\label{eqn:kappa}
\end{equation}
We use the projected potential $\Psi$, with the Laplacian operator $\nabla^2$ defined using the 
flat sky approximation as $\nabla^2\equiv \partial^2/{\partial\bm{\theta}^2}$ and $\chi$ is the comoving
distance (assuming a spatially flat universe). Note that $\chi$ is related to redshift $z$ via the relation 
$d\chi=dz/H(z)$, where $H(z)$ is the Hubble parameter at epoch $z$. The lensing efficiency function 
$W$ is given by
\begin{equation}
W(\chi)=\frac{3 \Omega_{m0} H_0^{2}}{2c^{2}} \frac{\chi}{a(\chi)}
\int\!\!d\chi_s~ n_s(\chi_s) \frac{\chi_{\rm s}-\chi}{\chi_s},
\label{eqn:weightgl}
\end{equation}
where $n_s(\chi_s)$ is the redshift selection function of source galaxies, $H_0$ is the Hubble 
constant today, $\Omega_{m0}$ is the matter density today, and $c$ is the speed of light. For 
simplicity, we take source galaxies to be at a single redshift $z_s$, so that 
$n_s(\chi)=\delta_D(\chi-\chi_s)$.  

Here we are interested in the deflection angle, which at a given point in the photon trajectory is 
given by the transverse gradient of the gravitational potential: $-2\nabla_\perp \phi$. The projected 
lensing potential $\Psi$ above is defined by
\begin{equation}
\Psi = -2\int_0^{\chi_s} \frac{\chi(\chi_s-\chi)} { \chi_s} \phi(\chi) d\chi .
\end{equation}
The net transverse deflection angle for a photon traveling from a source galaxy to the 
observer is then simply given by 
\begin{equation}
\vec{\alpha} = \vec{\nabla} \Psi.
\label{eqn:alpha}
\end{equation}
In Fourier space this gives 
\begin{equation}
\tilde\alpha_i(\ell) = -\ell_i \tilde\Psi = \frac{2\ell_i\tilde\kappa(\ell)}{\ell^2}.
\label{eqn:fourier_alpha}
\end{equation}
where $i=1,2$ corresponds to the x- and y-component of $\vec{\alpha}$. 
We can now write down the two-point correlation function of the deflection angles
\begin{equation}
\xi_{\alpha_i\alpha_j}(\theta)=\langle \alpha_i (\bm{\theta}_1)
 \alpha_j (\bm{\theta}_2)\rangle,
\label{eqn:shearcorrelation}
\end{equation}
with $\theta = |\bm{\theta}_1 - \bm{\theta}_2|$. We denote by $\xi_{\alpha\alpha}(\theta)$ 
the sum of the $x$- and $y$-components. 

The deflection angle power spectrum at angular wavenumber $\ell$ is the Fourier transform of 
\Eref{eqn:shearcorrelation}. Using the equations above we can express it in terms of the mass 
density power spectrum $P_\delta$: 
\begin{equation}
C_{\alpha\alpha}(\ell) = \frac{4 C_{\kappa\kappa}(\ell)}{\ell^2} = 
\frac{4}{\ell^{2}} \int_0^{\chi_s} d\chi \,{W(\chi)^2  \over \chi^2}
 P_\delta (\frac{\ell}{\chi}, \chi) .
\end{equation} 
The integral is dominated by the mass fluctuations at a distance about half-way to the source 
galaxies. The factor of $\ell^2$ in the denominator, compared to $C_{\kappa\kappa}$, leads 
displacements to be dominated by much larger scale modes than the convergence (or shear) fields. 

\Fref{fig:Cl_alpha} shows the predicted power spectrum and angular correlation function of $\alpha$ 
assuming a standard dark energy dominated, spatially flat cosmology: $h=0.7$, $\Omega_{m}=0.3$, 
$w=-0.95$, $\sigma_{8}=0.8$, for four different source redshifts. 
Similar to $C_{\kappa\kappa}$, the amplitude is larger at high redshift, as there is more integrated 
matter causing the deflection. As noted above, the power spectrum for the deflection angles peaks 
at lower $\ell$. We discuss below the observational consequences of this feature.

\section{Observational implication}
\label{sec:observation}

\begin{figure*}
\begin{center}
  \includegraphics[scale=0.5]{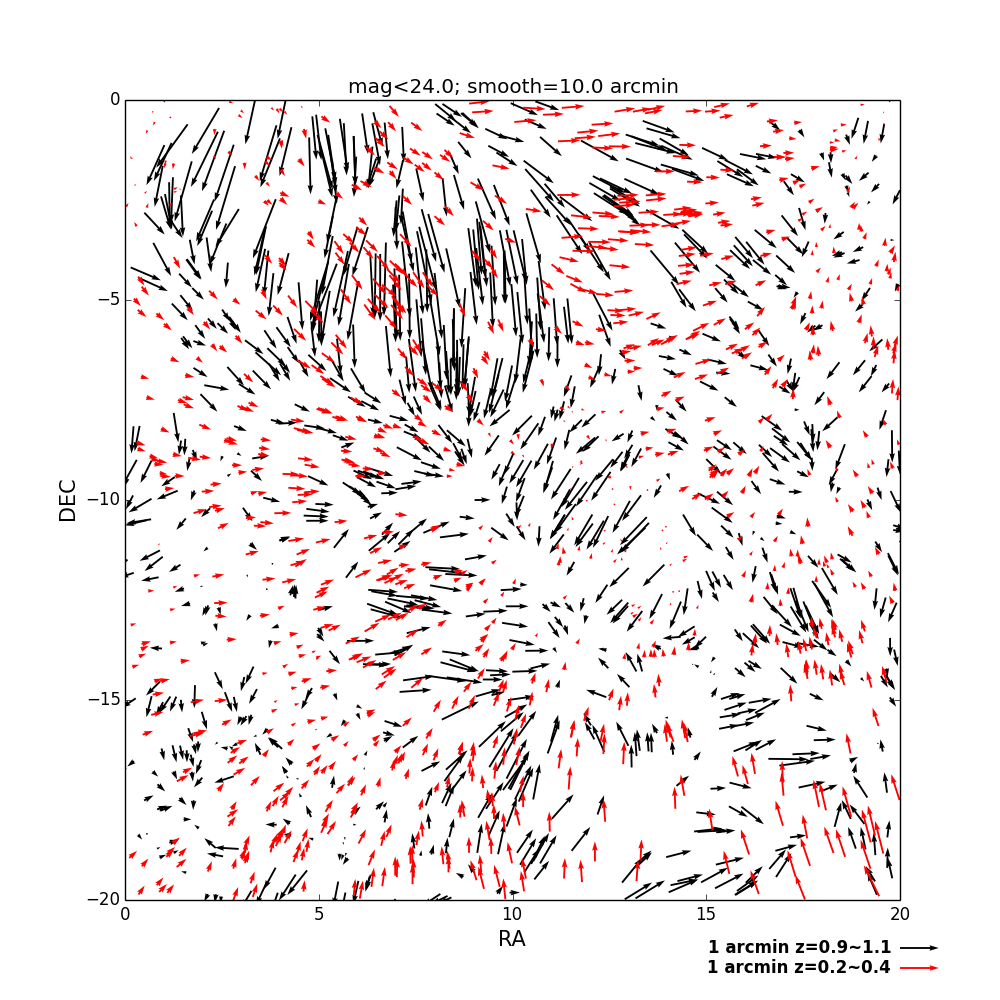}
  \caption{ The displacement of galaxies is shown on a $20\times 20$ deg simulated patch. The arrows 
  have a separate scale shown with the 1 arcmin arrow on the top right. They show the displacement due to 
  lensing of source galaxies at two different redshifts: 0.3 in red and 1.0 in black. Note that the scale for the 
  deflections is different from the scale on the x- and y-axis that shows the field of view. The displacements 
  are fairly coherent on degree scales, but galaxies at different redshifts move relative to each other by up to 
  an arcminute. So if we delens, say, a Hubble Deep Field sized patch of the sky, the galaxy locations will be 
  scrambled! }
  \label{fig:alpha_arrow}
\end{center}
\end{figure*}

We can now estimate the numerical values of the deflection angles and investigate implications. As discussed 
above, the peak of the power spectrum at $\ell < 100$ means that the deflections are coherent over 
several degree-sized scales. Photons from source galaxies at redshift  $\sim 1$, with co-moving distance of a few 
Gpc, are deflected by several tens of independent structures along their path to the observer. The physical 
picture is similar to a random walk -- the further the source is, the more steps the photon will take in the random 
walk, therefore a larger deflection angle. The amplitude of the deflection, according to the correlation function in 
\Fref{fig:Cl_alpha}, is of order 1 arcminute at redshift $\sim1$. So galaxies at different redshifts are displaced 
by fractions of an arcminute, many times the typical separation of galaxies. 
\textit{This suggests that, if one were to effectively ``delens'' the galaxy positions on the sky, it will visibly shuffle 
the galaxies near each other on the sky but at different redshifts.} 

The lensing deflection and its redshift dependence can be visualised in simulations as shown in \Fref{fig:alpha_arrow}. 
Details of the simulations are described in \Sref{sec:sims}. \Fref{fig:alpha_arrow} provides the visual picture of 
the deflection effect in observational data -- galaxies at different redshifts are shuffled around at a  significant 
level. Objects at high redshift ($z\sim1$) are typically deflected more relative to the low-redshift ($z\sim0.3$) 
objects, with the difference being of order an arcminute. The full field of the simulation is 20$\times$20 deg$^{2}$; 
we can see that the deflections are coherent on degree-size patches. 

The effect of this deflection field on a generic cosmological correlation function has been studied in 
\citet{2008PhRvD..78d3508D} and references therein. In particular, they pointed out that for galaxy correlation functions, the effect  
of the deflection field is generally at the percent level. 
Lensing deflections smooth out oscillating features in the correlation function (e. g. the BAO peak) just as they do for the CMB angular power spectrum.  
 
In the next section we investigate the effect of the deflection field on one of the cosmological measurements 
that does not fall into the category of the general correlation functions studied in \citet{2008PhRvD..78d3508D}. 
This measurement is especially relevant as it depends on correlating galaxies that are well separated in redshift 
-- these galaxies are thus affected at a very different level by the deflection field as can be seen in 
\Fref{fig:alpha_arrow}. 
 
\section{Galaxy-galaxy lensing}
\label{sec:g-g}

\begin{figure}
\begin{center}
    \includegraphics[scale=0.16]{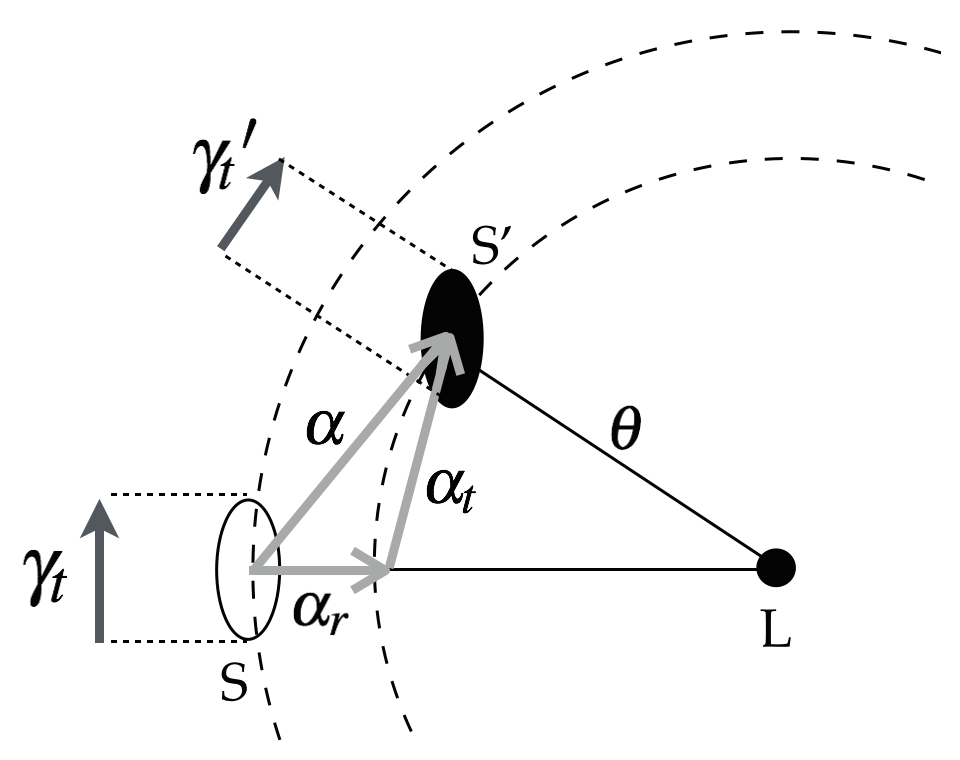}
  \caption{Illustration of the impact on galaxy-galaxy lensing from deflections between the lens and observer. The solid black ellipse 
  (S') represents the observed source galaxy whereas the hollowed ellipse (S) represents the true galaxy position. The 
  solid circle (L) is the foreground lens galaxy. The deflection angle $\alpha$  can be decomposed  
  into a radial ($\alpha_{r}$) and a tangential ($\alpha_{t}$) component with respect to the lens galaxy. $\alpha_{r}$ 
  affects the relation between observed angular separation and transverse distance between source and lens: $\chi \theta \rightarrow \chi (\alpha_{r}+\theta)$. $\alpha_{t}$ affects the measured 
  tangential shear signal. The true signal $\gamma_{t}$, indicated by the dark-grey arrow, is diluted to 
  ${\gamma_{t}}^{\prime}$ because the apparent tangential direction has changed when the galaxy image is displaced.}
  \label{fig:cartoon}
\end{center}
\end{figure}

Galaxy-galaxy lensing measures the cross-correlation of foreground galaxy positions with background galaxy 
shears. The cosmological signal can be measured by averaging the tangential component of the background 
galaxy ellipticities in circular annuli centered on the foreground galaxy, denoted as $\langle\gamma_t\rangle(\theta)$. 
$\langle\gamma_t\rangle(\theta)$ is directly translated into the projected mass density through 
$\Sigma(R) = \Sigma_{\rm crit} \gamma_t(\theta)$ (we drop the averaging notation here on). Deflections between the 
lens and observer can perturb the relative positions of the lens and source galaxy. 
In \Fref{fig:cartoon} we illustrate both effects for one source-lens pair used to measure galaxy-galaxy lensing. The 
deflection angle $\vec{\alpha}$ is decomposed into a radial ($\alpha_{r}$) and a tangential component ($\alpha_{t}$). 
The figure shows how the two components introduce errors in the apparent source-lens distance and the measured 
tangential shear respectively.  

\begin{figure}
\begin{center}
  \includegraphics[scale=0.45]{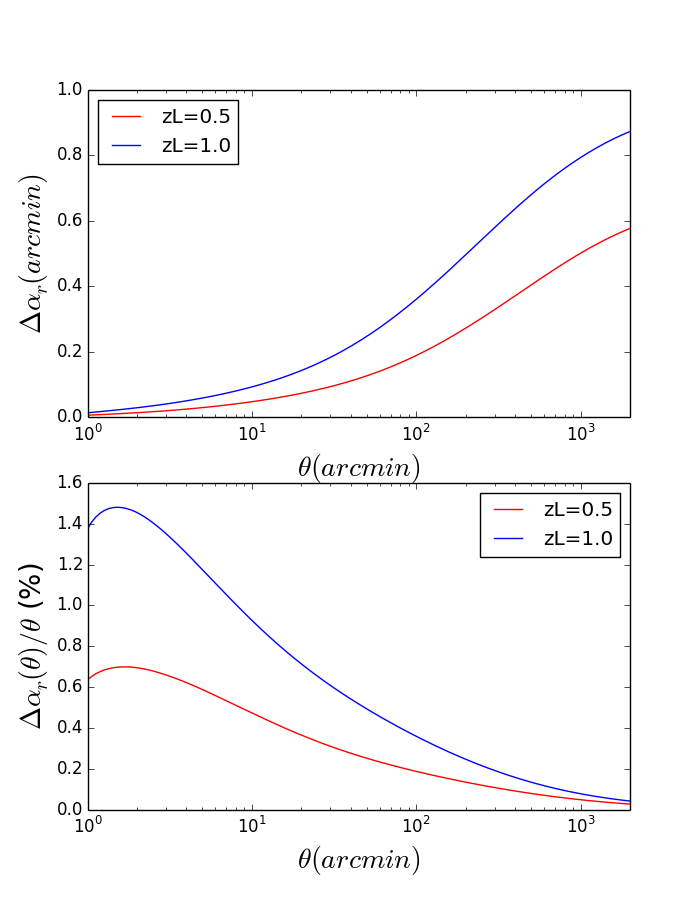}
  \caption{The radial component of the rms deflection angle $\Delta\alpha_{r}(\theta)$ vs. $\theta$ is shown in the top panel for two lens redshift 
  $z_{L}=$0.5 and 1.0. The ratio $\Delta\alpha_{r}/\theta$ shown in the bottom panel is relevant for the observable effects discussed in the text. 
}
  \label{fig:delta_alpha}
\end{center}
\end{figure}

For the radial deflection, we are  interested in the mapping 
from measured angle to inferred spatial separation on the lens plane. We are interested in the rms quantity 
$\Delta\alpha_{r}(\theta)$, the uncertainty induced in the angular position of the source galaxy relative to the lens in 
the radial direction which smears out features in the $\langle\gamma_t\rangle(\theta)$ curve. It is given by: 
\begin{equation}
\Delta\alpha_{r}(\theta)^2 \equiv \frac{1}{2} \langle (\vec{\alpha}(0) - \vec{\alpha}(\theta))^2\rangle =  \xi_{\alpha \alpha}(0)
 - \xi_{\alpha \alpha}(\theta). 
\label{eq:delta_alpha}
\end{equation}
The factor of 1/2 accounts for the fact that we are concerned with the radial component of $\Delta \vec{\alpha}$. 
$\Delta\alpha_{r}(\theta)$ has the effect of smearing the relation between observed angle $\theta$ and the distance 
from lens, i.e. 
\begin{equation}
R=\chi\theta \rightarrow R=\chi(\theta\pm\Delta\alpha_{r}).
\end{equation}
The mass profile $\Sigma(R)$ is then altered. We can estimate the effect by integrating over the foreground deflections, 
i.e. from $z=0$ to the lens redshift $z_L$. 
%Note that this effect arises due to the component of the deflection along the 
%radial direction from the lens galaxy. 
\Fref{fig:delta_alpha} shows the calculated $\Delta\alpha_{r}(\theta)$ for lenses at redshift 0.5 and 1.0, along with the 
fractional error of the lens-source distance used in galaxy-galaxy lensing. At lens redshift $z_L=1$ and $\theta=10$ 
arcminutes, the error on the spatial separation between the source and the lens is about 1\%. Depending on the shape 
of the $\langle\gamma_t\rangle(\theta)$ curve, this effect can introduce errors at different levels. It is only second order in $\Delta\alpha_{r}$, so unless the curvature is large, the effect is negligible. However, for 
particular cases such as satellite galaxy-galaxy lensing studied in \citet{2013MNRAS.430.3359L}, there can be sharp 
features in the curves, which would be smoothed out  due to the deflection effect.  Even for such cases we do not expect the effect to be a worry for ongoing surveys. 

A second effect arises due to the deflection in the tangential direction: it lowers the estimated tangential shear  
$\langle\gamma_t\rangle(\theta)$ because the tangent direction itself is incorrect due to the deflections that occur 
between the source and the lens. 
Imagine a circle centered on the lens galaxy, the source galaxy position along that circle is changed by a differential  
deflection between lens and source. We choose the tangential direction based on its observed 
position, while the lens galaxy's shear is tangent to the location where the light rays pass the lens plane. Unlike the radial effect, this does not depend on the shape of the $\langle\gamma_t\rangle(\theta)$ curve, but it is also a second order effect and is therefore well below a percent (note that the contribution to the ``cross'' component of the shear, usually used as a diagnostic of systematics, is first order and can be larger than a percent). 
 Hence both effects may be relevant only for LSST and Euclid-era surveys that will achieve 
sub-percent statistical errors on the galaxy-galaxy lensing measurement.

\section{Reconstruction of the deflection angle field}
\label{sec:reconstruction}

Consider now the possibility of reconstructing the galaxies' true positions from observable quantities, 
which we will call ``delensing''. We present two approaches to delensing that differ in how the convergence 
field $\kappa$ is obtained. 
The first method uses weak lensing shear measurements to construct the convergence field. The 
second uses the foreground galaxy density field with known galaxy bias to infer the convergence -- this 
method is used if the shear-based convergence map is unavailable. The relation between the convergence 
and the lensing potential in Fourier space then follows from \Eref{eqn:kappa} and \Eref{eqn:alpha}.

The fundamental limit of both methods is the finite size of the observed field. Since the reconstruction is a 
non-local operation, we cannot recover the full information with only limited Fourier modes inside the field. 
To first order, if we want to capture the dominant contributions to the deflection field, a field that spans linear 
scales  $\sim$10 degrees ($\ell \sim 10$) or larger is required. As a result, shear-based reconstruction has 
only become feasible with ongoing lensing surveys such as DES. (The Canada-France Hawaii Telescope 
Lensing Survey covers $\sim170$ deg$^{2}$ area, but the coverage is not contiguous.) Also note that 
operationally, constructing a continuous field from discrete galaxy measurements in both cases require 
smoothing and/or pixelization. As a result, there is also inevitable information loss on small scales, though 
this is not an issue in practice as we discuss below. 

\subsection{Reconstruction from weak lensing shear}
\label{sec:alpha_from_epsilon}

The convergence can be calculated from the measured shear, which is a noisy but observable quantity. 
We use the method proposed by \citet{1993ApJ...404..441K} to construct the continuous convergence 
map from shear. The Fourier transform of the observed shear relates to the convergence through 
\begin{equation}
\kappa(\ell) - \kappa_0= D^*(\ell) \gamma(\ell),
\label{eq:ks_ft}
\end{equation}
where $\kappa_0$ is the average projected mass, i.e. $\kappa$ for $\ell=0$, and 
\begin{equation}
D(\ell) = \frac{\ell_1^2 - \ell_2^2 + 2 i \ell_1 \ell_2}{|\ell|^2}.
\label{eq:D}
\end{equation}

The major source of error in this approach comes from the large scatter in the galaxies' intrinsic shape 
(shape noise), which contributes to a random error in the shear and therefore the $\kappa$ estimates. 
Since the deflection angle field is dominated by scales larger than a degree, the number density of 
galaxies is normally sufficient to ensure that shape noise smoothed on these scales is smaller than the 
signal. It is however not negligible for current surveys. 
Practically, the dominance of large angular scales also suggests that smoothing $\kappa$ to suppress 
the noise will not affect the inferred deflection angles. 

\subsection{Reconstruction from the galaxy density field}
\label{sec:alpha_from_deltag}

Alternatively, one can estimate the convergence field through the foreground galaxy density field if we 
assume some model for the galaxy bias $b$, so that $\delta_{g}= b\delta_{m}$, where $\delta_{g}$ is the 
over-density in the galaxy number counts and $\delta_{m}$ is the mass over-density. Note that in principle, 
$b$ could be a function of redshift and other galaxy properties. Using the linear bias relation, we can 
re-write the equation for $\kappa$, 
\Eref{eqn:kappa}, as:  
\begin{equation}
\kappa (\bm{\theta}) \approx \int\!\!d\chi W(\chi) b^{-1} \delta_{g}[\chi, \chi\bm{\theta}].
\label{eqn:kappag}
\end{equation}

The main advantage of this approach in constructing $\kappa$ is that it by-passes the shape noise, and 
does not involve the Fourier transform in \Eref{eq:ks_ft} that introduces numerical errors in practice. An 
independent estimate of galaxy bias over the range of lens redshift, however, is needed for this method. 

\begin{figure*}
\begin{center}
  \includegraphics[width=0.49\textwidth]{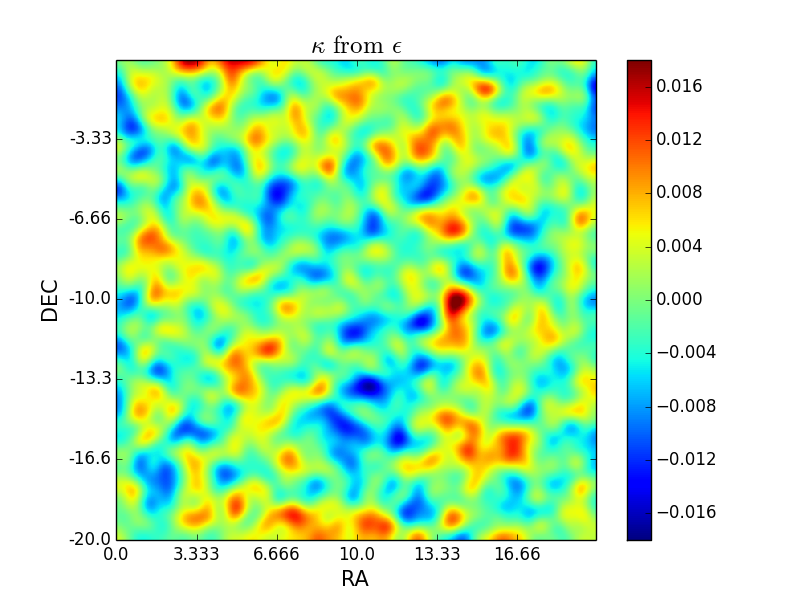}
  \includegraphics[width=0.49\textwidth]{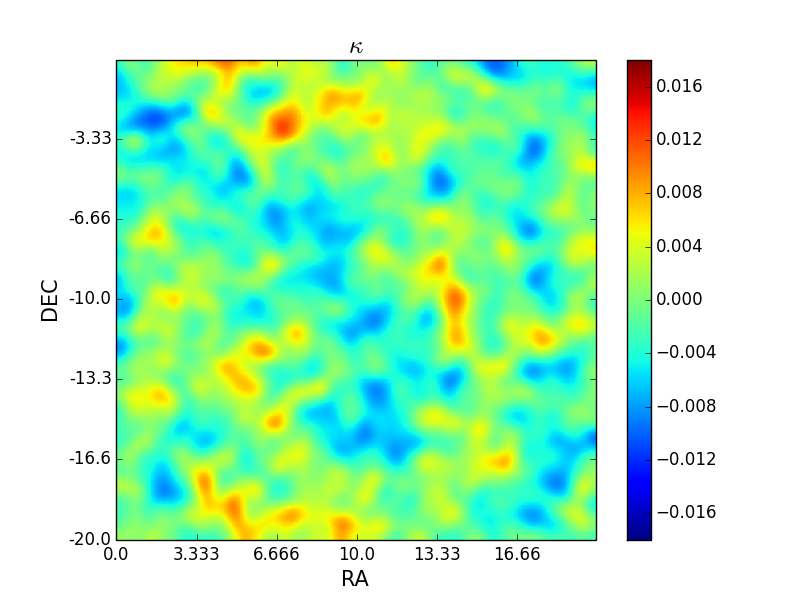}
   \includegraphics[width=0.49\textwidth]{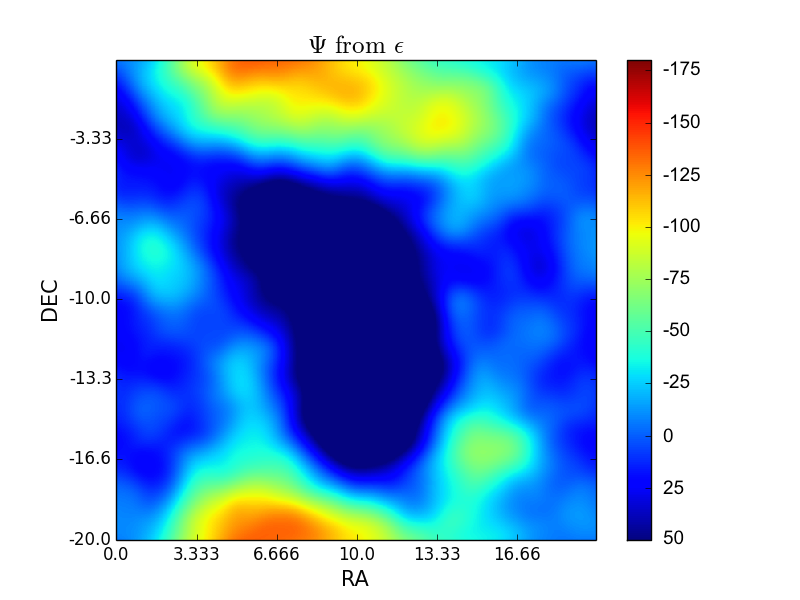}
   \includegraphics[width=0.49\textwidth]{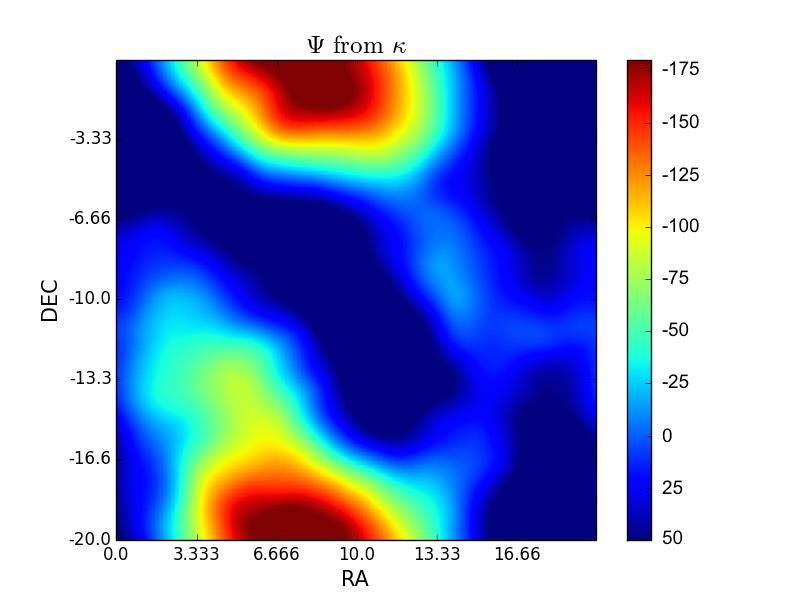}
   \includegraphics[width=0.42\textwidth]{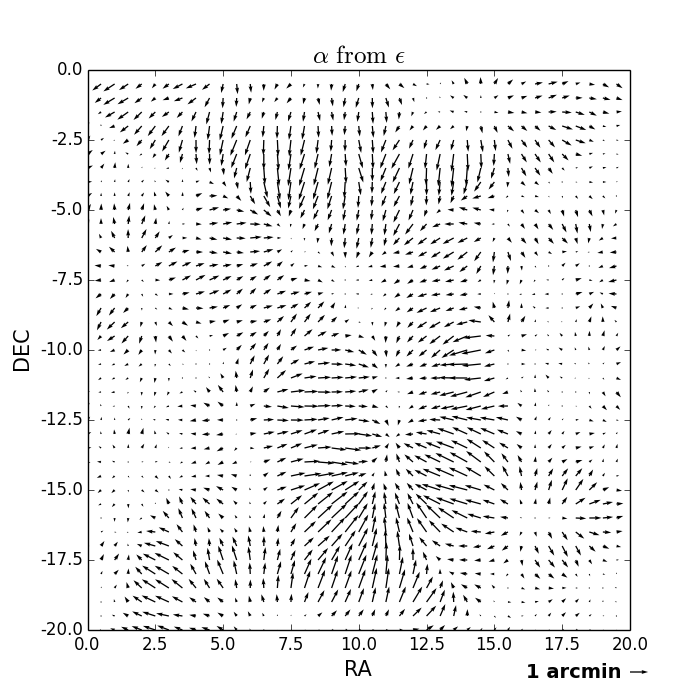} \hspace{0.4in}
   \includegraphics[width=0.42\textwidth]{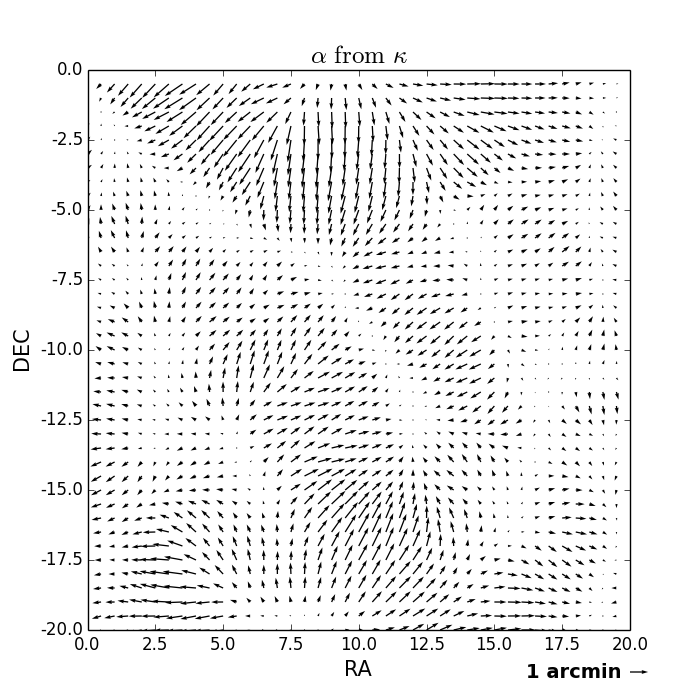}  \hspace{0.3in}
  \caption{Maps generated for a 20$\times$20 deg$^{2}$ field, in the redshift range $0.9<z<1.1$. All maps are generated 
  on 2$\times$2 arcminute$^{2}$ pixels with a 20 arcminute RMS Gaussian smoothing. Left: $\kappa$ (top), 
  $\Psi$ (middle), and $\vec{\alpha}$ (bottom) maps generated from galaxy shape ($\epsilon$) measurements. These maps 
  are generated with a galaxy sample with magnitude cut $i<24.0$, which is realistic for future wide-field surveys. Right: 
  the same maps with the true $\kappa$. The difference between left and right shows the effect of shape noise as well as 
  the $\kappa$ reconstruction from galaxy shapes. 
  }
  \label{fig:kappa_psi_maps}
\end{center}
\end{figure*}

\begin{figure*}
\begin{center}
  \includegraphics[width=0.49\textwidth]{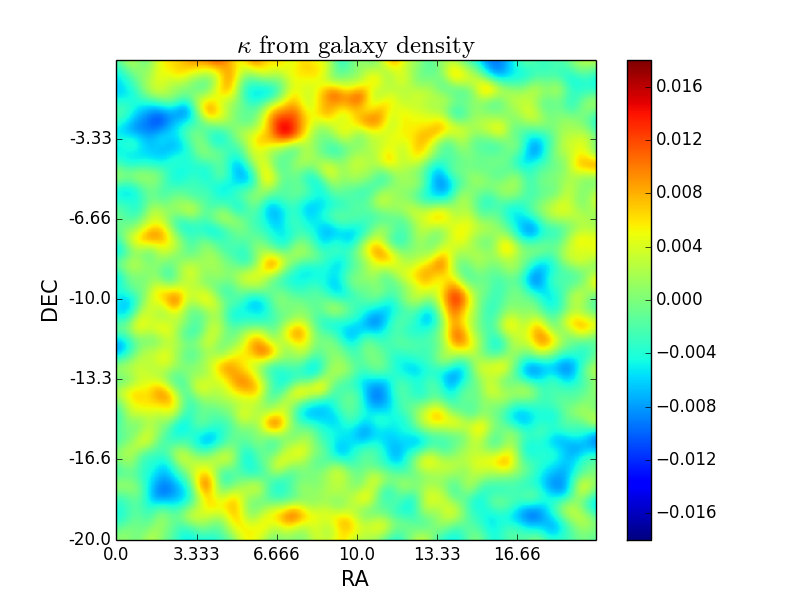} 
  \includegraphics[width=0.38\textwidth]{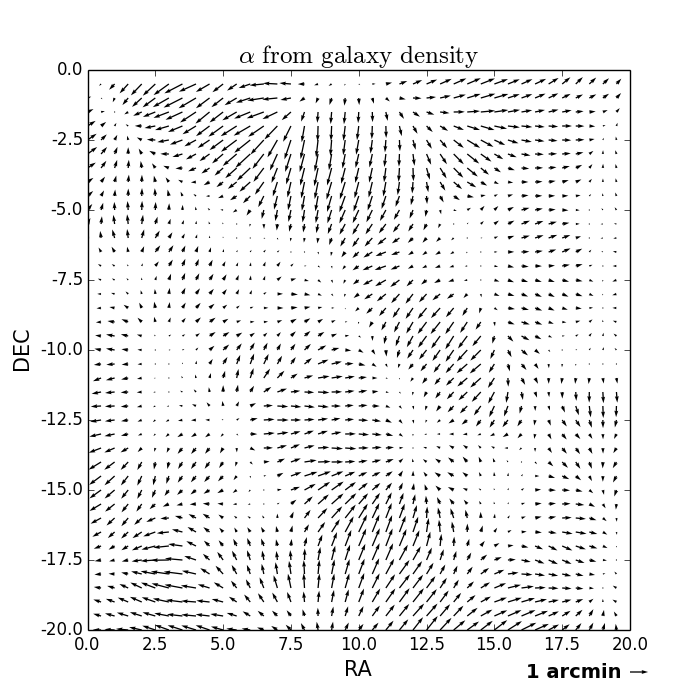} 
    \caption{The $\kappa$ (left) and $\vec{\alpha}$ (right) field generated from the foreground galaxy density field, 
    assuming a constant galaxy bias $b=2$. The reconstructed $\vec{\alpha}$ field is almost identical to that 
    constructed from the true $\kappa$. This result depends on the simplicity and knowledge of galaxy bias.  }
  \label{fig:kappag}
\end{center}
\end{figure*}

\subsection{Delensing  simulations}
\label{sec:sims}

\begin{figure*}
\begin{center}
   \includegraphics[scale=0.3]{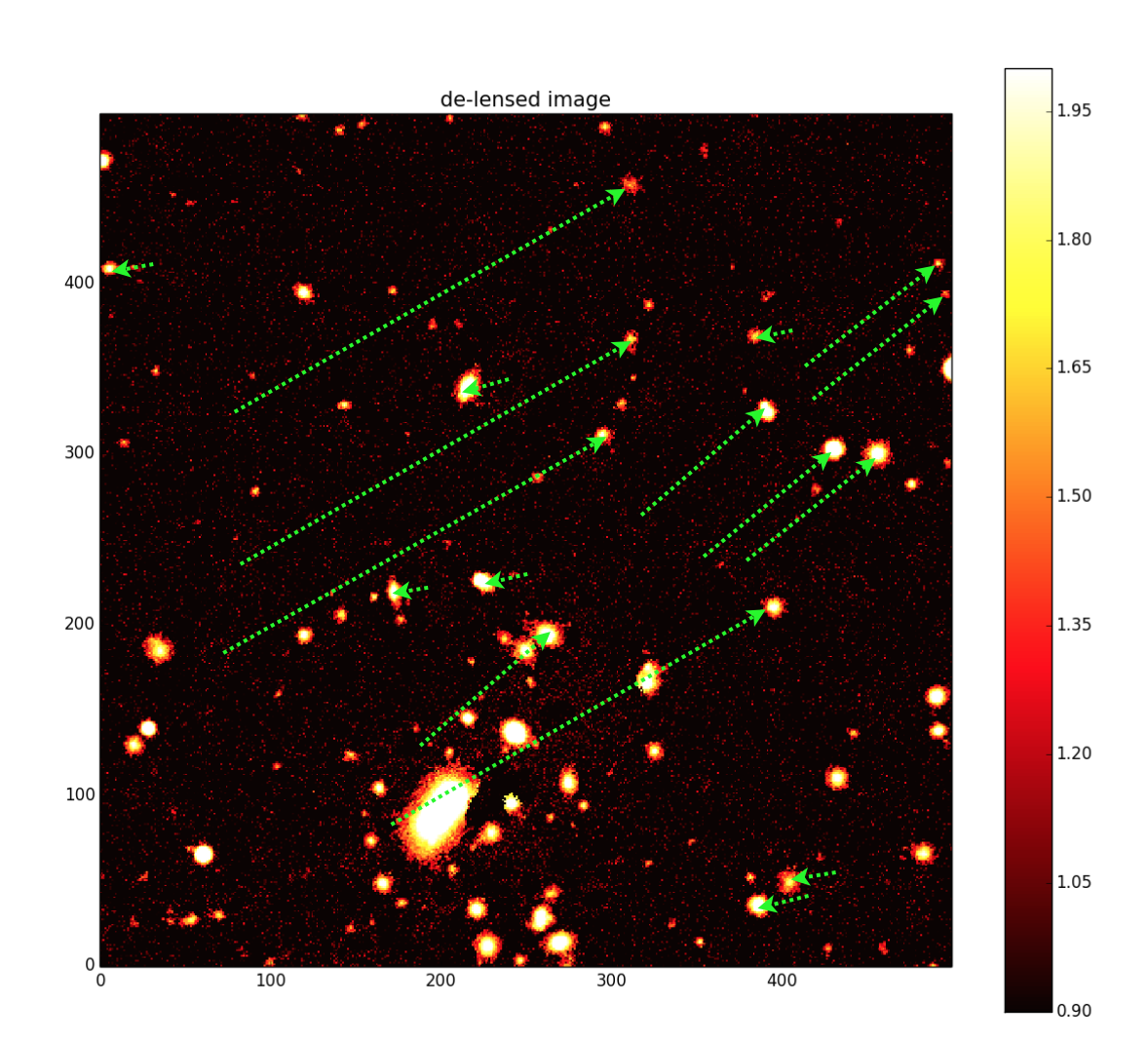} 
    \caption{The delensed field (shown in detail below in \Fref{fig:remapping}) with arrows indicating the displacement of 
    some of the galaxies from the observed image to the de-lensed image. The galaxy position in 
    the observed image is at the beginning of the arrow and it moves to the end-point of the arrow in  
    the delensed image. The image size is 2.25$\times$2.25 arcminute$^{2}$ area. Visually one can see 
    that there are three different lengths and orientations of the arrows in this image, associated with the 
    three redshift bins shown in \Fref{fig:remapping}. With tomographic lensing mass maps, displacement fields over a wide range of source redshifts can be obtained from wide field surveys. }
  \label{fig:remapping_zoom}
\end{center}
\end{figure*}

A series of tests with simulations is shown here to demonstrate 
the operational steps in performing the reconstruction. We use the mock galaxy catalog generated by the 
algorithm Adding Density Determined Galaxies to Lightcone Simulations (ADDGALS; Wechsler et al, in 
preparation; Busha et al, in preparation). The lensing information (deflection angle, convergence and shear) 
is included for each galaxy according to a ray-tracing procedure described in \citet{2013MNRAS.435..115B}. 
The total area of the full simulation covers a quarter of the sky. The lensing $\kappa$ maps are 
accurate down to about an arcminute, which is sufficient for our purposes. We use a 20$\times$20 
deg$^{2}$ patch of the simulation to demonstrate the principle of reconstruction.
The following parameters for each galaxy in the catalog are used: true position, lensed position, 
magnitude, ellipticity $\epsilon$, convergence $\kappa$, and shear $\gamma$ at the lensed position. 
Previously we have shown in \Fref{fig:alpha_arrow} an example of the deflection angles of galaxies for two 
different redshifts from the simulations.  

We first demonstrate the reconstruction procedure described in \Sref{sec:alpha_from_epsilon}. We use the 
high-redshift galaxy sample (0.9$<$z$<$1.1) in \Fref{fig:alpha_arrow} with a magnitude cut of $i<24$ to simulate 
what is expected in a magnitude-limited dataset. In the 20$\times$20 deg$^{2}$ area, there are $\sim$2.68/arcmin$^{2}$ 
galaxies in this sample. The median redshift is $\sim1.0$. We pixelate the $\epsilon$ values into a 600$\times$600 
grid, with each pixel being 2$\times$2 arcmin$^{2}$. The rms of the ellipticity distribution per component (i.e. shape 
noise per shear component) is $\sim0.27$ at this magnitude cut. We then apply a 20 arcminute rms Gaussian smoothing to the 
maps to suppress noise \citep{1993ApJ...404..441K}. The $\epsilon$ maps are converted to $\kappa$ maps according 
to \Eref{eq:ks_ft}. From the $\kappa$ maps we reconstruct the lensing potential $\Psi$ according to \Eref{eqn:kappa}. 
And finally we calculate the $\vec{\alpha}$ field using \Eref{eqn:fourier_alpha}. The left panel of 
\Fref{fig:kappa_psi_maps} shows the series of maps generated in this procedure. The right panel of 
\Fref{fig:kappa_psi_maps} shows the same series of maps but using the true $\kappa$ instead of using $\epsilon$ to 
make the $\kappa$ maps. The difference in the two panels comes from shape noise as well as errors introduced in 
\Eref{eq:ks_ft}.
The median error in the reconstructed $\vec{\alpha}$ compared to the true $\vec{\alpha}$ in the simulations is 
$\sim 30$ degrees in the direction angle of and $50\%$ in the magnitude of $\vec{\alpha}$. The 
reconstruction degrades when reducing the area of the simulation used. With a factor of $\sim$8 smaller area, 
the reconstruction is essentially consistent with noise. This confirms that most of the power in the deflection field 
comes from the large scales. Several other sources of intrinsic errors are expected in this reconstruction:
\begin{itemize}
\item We have implicitly assumed the average $\kappa$ is approximately the same as the $\kappa$ at the 
median redshift of the sample.  
\item The galaxy intrinsic shape introduces noise in the $\kappa$ maps and therefore the reconstruction.
\item Numerical errors in the simulation, interpolation and the flat sky approximation -- these are expected to 
be negligible. 
\end{itemize}

Next, we demonstrate the reconstruction procedure based on the foreground galaxy distribution, described in 
\Sref{sec:alpha_from_deltag}. We construct the $\kappa$ map using galaxies from $z=0$ to $z=1$, weighted 
by the lensing kernel $W(\chi)$ (\Eref{eqn:weightgl}). We assume a constant bias $b=2.0$ and a magnitude cut 
of $i<24$. The $\Psi$ and $\vec{\alpha}$ maps are then obtained as before. The $\kappa$ and the final $\vec{\alpha}$ 
maps from this approach are shown in \Fref{fig:kappag}. We find that the $\vec{\alpha}$ map reconstructed in this 
fashion is much less noisy given that there is no scatter from associated with shape noise. The rms error 
between the $\kappa$ map generated from the ellipticities and the true $\kappa$ map in our test is 
$\sim$0.0036, whereas the rms error for the $\kappa$ map generated from the galaxy density is $\sim0.0009$. 
However, the method requires a priori knowledge of galaxy bias before performing the reconstruction. The fact 
that we get good results is likely due to the simplicity of the galaxy assignment in the simulations. 

One possible improvement to the two reconstruction methods discussed above is to combine the two, similar 
to \citet{2012MNRAS.424..553A} and use the information in the density field as well as the shear field to 
constrain the deflection and the bias simultaneously. Note that we only need an estimate of the average bias 
on large scales over the lens redshift range, thus the bias is needed to less accuracy compared to that used in 
\citet{2012MNRAS.424..553A}.

\section{Conclusion}
\label{sec:conclusion}

We have studied   lensing deflections in the context of ongoing and planned wide-field 
galaxy surveys. The power spectrum of the deflection field peaks at  large angular scales, so the deflections are 
coherent over degree-sized scales.  The rms deflection is about one arcminute at redshift $\sim 1$. The 
redshift-dependence of the deflection field implies that galaxies that are nearby on the sky can be displaced by 
fractions of an arcminute relative to each other. So measurements 
requiring cross-correlation of galaxies at different redshift can be affected. We discuss galaxy-galaxy 
lensing and estimate that the error associated with ignoring the deflection field is at the sub-percent level and therefore not a concern for ongoing lensing surveys.  It may be worth examining further for particular galaxy lens populations  for LSST and Euclid-era surveys. 

The main purpose of the paper is to undo the effect of deflections on wide field images by ``delensing'' galaxy positions. We implement two approaches to delensing: one is based on mass maps that use the measured lensing shear, while the second uses the galaxy density field as a proxy for the lensing mass distribution. We show that a field of several hundred square degrees is required for the reconstruction to be  
reasonably accurate. We expect better reconstructions with even larger fields.  Ongoing imaging surveys such as DES  will for the first time have the capability to delens the galaxy distribution. 

Once the deflection field is reconstructed, one can delens the pixel-level images via standard 
image-remapping programs (e.g. OpenCV\footnote{\url{http://docs.opencv.org/index.html}}). In 
\Fref{fig:remapping_zoom} we demonstrate an example of delensing the simulated sky by remapping every pixel 
in the image. We bin the galaxies into 4 redshift bins, and delens each bin by the  deflection field at the mean 
redshift of that bin. Noise and stars are not delensed. For visualization purposes, we zoom into a 2.25$\times$2.25 
arcminute$^{2}$ area (500$\times$500 pixels in our simulated image). We can see a clear change in the galaxy 
apparent positions on the sky. More details of the image generation are described in \Aref{sec:warp}.  Tomographic lensing mass maps from wide field surveys will allow us to reconstruct the entire trajectories of lights rays from distant galaxies to us. 

\appendix
\section{Pixel-level delensing}
\label{sec:warp}

\begin{figure*}
\begin{center}
 \includegraphics[scale=0.4]{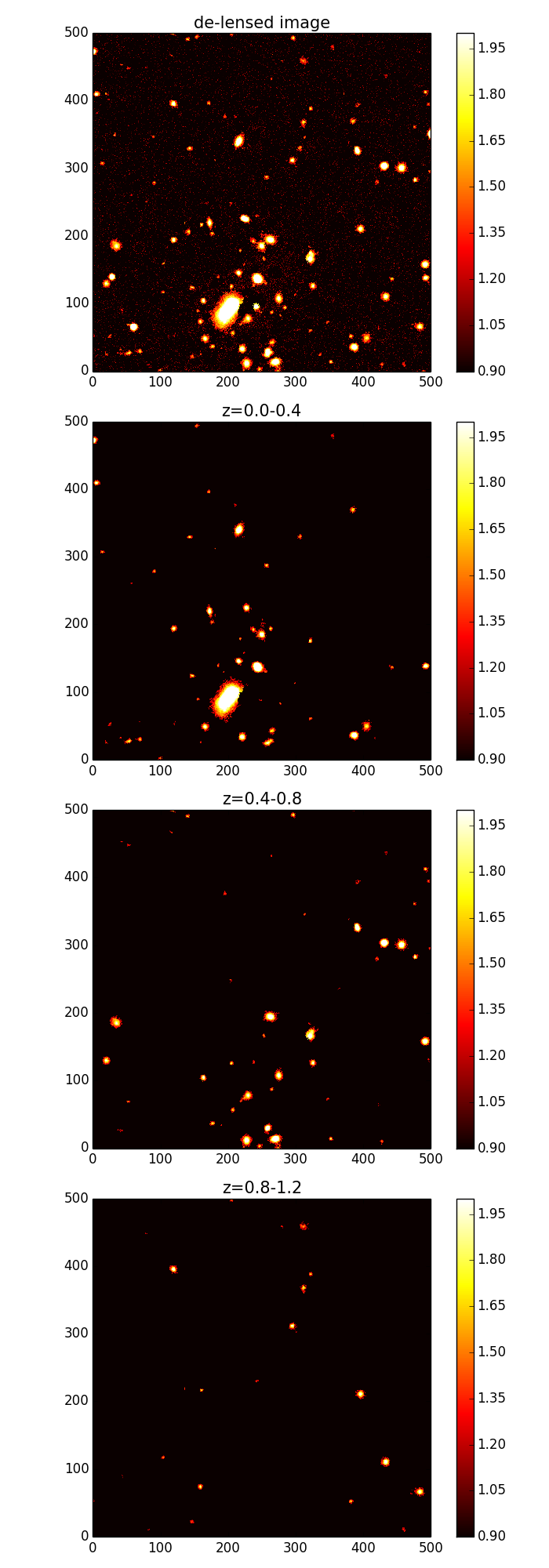} 
  \includegraphics[scale=0.4]{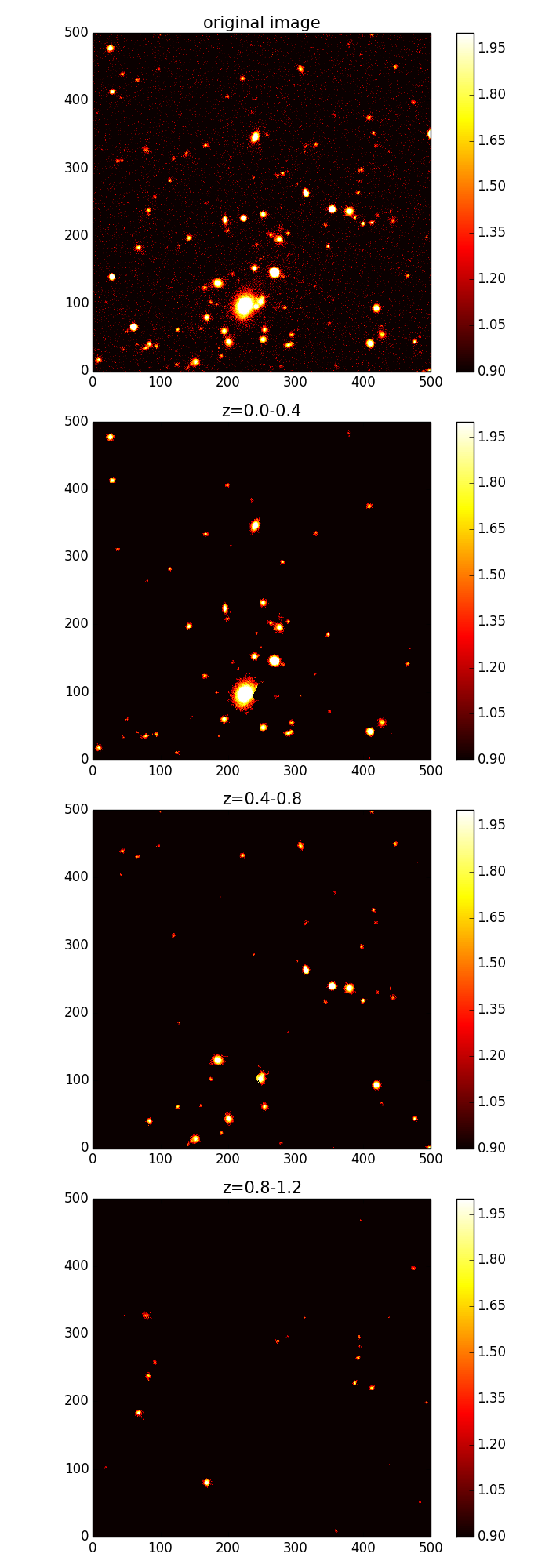} \\ 
    \caption{An example of pixel-level delensing with simulations. The observed images on the left 
    is delensed into the right-hand-side images. The top panel is the full image (same as 
    \Fref{fig:remapping_zoom}), which is a sum of the different redshift components. We show 
    three redshift bins in the bottom three rows, which correspond to the three coherent fields in 
    \Fref{fig:remapping_zoom} that have different lengths of arrows. Visually we can see the 
    high-redshift bins are shifted the most, and the relative relation between objects are changed 
    significantly before and after lensing. Note also that the $0.0 < z < 0.4$ bin shifts in an opposite 
    direction compared to the others. All images are 500$\times$500 pixels, with pixel scale is 0.27", 
    which yields a 2.25$\times$2.25 arcminute$^{2}$ area. The image colours are in logarithmic 
    scales.
    }
  \label{fig:remapping}
\end{center}
\end{figure*}

We perform a pixel-level delensing using images simulated by the Ultra Fast Image Generator 
\citep{2013A&C.....1...23B}. Pixels associated with each object are identified with the software 
Source Extractor \citep{1996A&AS..117..393B}, then remapped according to the displacement field 
at that redshift using the OpenCV software. There are visible image artefacts associated with the 
fact that we can only delens on individual pixels in this approach, i.e. thus blended objects cannot 
be individually delensed. The upper panel of \Fref{fig:remapping} shows the original image as in 
\Fref{fig:remapping_zoom} and the lower panels show how we decompose the image into the 
different redshift components.  
 
\vspace{0.2in}
 {\it Acknowledgement. } We acknowledge helpful
discussions with Adam Amara, Gary Bernstein, Sarah Bridle, Joseph Clampitt, Eric Huff, Elisabeth Krause and Mike Jarvis. 
We are grateful to Vinu Vikram for discussions and related collaborative work.  
We thank the simulation team, especially Matt Becker, Michael Busha and Risa Wechsler, 
responsible for the mock catalogs used in our tests and Lukas Gamper and Joel Berge for the 
image simulations. This work is supported in part by the Department of Energy grant 
DE-SC0007901 and the Swiss National Science Foundation grant 200021-149442.

\bibliography{Deflection_v6.bbl}

\begin{thebibliography}{16}
\expandafter\ifx\csname natexlab\endcsname\relax\def\natexlab#1{#1}\fi

\bibitem[{{Amara} {et~al}\mbox{.}(2012){Amara}, {Lilly}, {Kova{\v c}},
  {Rhodes}, {Massey}, {Zamorani}, {Carollo}, {Contini}, {Kneib}, {Le Fevre},
  {Mainieri}, {Renzini}, {Scodeggio}, {Bardelli}, {Bolzonella}, {Bongiorno},
  {Caputi}, {Cucciati}, {de la Torre}, {de Ravel}, {Franzetti}, {Garilli},
  {Iovino}, {Kampczyk}, {Knobel}, {Lamareille}, {Le Borgne}, {Le Brun},
  {Maier}, {Mignoli}, {Pello}, {Peng}, {Montero}, {Presotto}, {Silverman},
  {Tanaka}, {Tasca}, {Tresse}, {Vergani}, {Zucca}, {Barnes}, {Bordoloi},
  {Cappi}, {Cimatti}, {Coppa}, {Koekoemoer}, {L{\'o}pez-Sanjuan}, {McCracken},
  {Moresco}, {Nair}, {Pozzetti}, \& {Welikala}}]{2012MNRAS.424..553A}
{Amara} A. {et~al.}, 2012, \mnras, 424, 553

\bibitem[{{Bartelmann} \& {Schneider}(2001)}]{2001PhR...340..291B}
{Bartelmann} M., {Schneider} P., 2001, \physrep, 340, 291

\bibitem[{{Becker}(2013)}]{2013MNRAS.435..115B}
{Becker} M.~R., 2013, \mnras, 435, 115

\bibitem[{{Berg{\'e}} {et~al}\mbox{.}(2013){Berg{\'e}}, {Gamper},
  {R{\'e}fr{\'e}gier}, \& {Amara}}]{2013A&C.....1...23B}
{Berg{\'e}} J., {Gamper} L., {R{\'e}fr{\'e}gier} A., {Amara} A., 2013,
  Astronomy and Computing, 1, 23

\bibitem[{{Bertin} \& {Arnouts}(1996)}]{1996A&AS..117..393B}
{Bertin} E., {Arnouts} S., 1996, \aaps, 117, 393

\bibitem[{{Dodelson}, {Schmidt} \& {Vallinotto}(2008){Dodelson}, {Schmidt}, \&
  {Vallinotto}}]{2008PhRvD..78d3508D}
{Dodelson} S., {Schmidt} F., {Vallinotto} A., 2008, \prd, 78, 043508

\bibitem[{{Heymans} {et~al}\mbox{.}(2012){Heymans}, {Van Waerbeke}, {Miller},
  {Erben}, {Hildebrandt}, {Hoekstra}, {Kitching}, {Mellier}, {Simon},
  {Bonnett}, {Coupon}, {Fu}, {Harnois D{\'e}raps}, {Hudson}, {Kilbinger},
  {Kuijken}, {Rowe}, {Schrabback}, {Semboloni}, {van Uitert}, {Vafaei}, \&
  {Velander}}]{2012MNRAS.427..146H}
{Heymans} C. {et~al.}, 2012, \mnras, 427, 146

\bibitem[{{Hoekstra} \& {Jain}(2008)}]{2008ARNPS..58...99H}
{Hoekstra} H., {Jain} B., 2008, Annual Review of Nuclear and Particle Science,
  58, 99

\bibitem[{{Huff} \& {Graves}(2011)}]{2011arXiv1111.1070H}
{Huff} E.~M., {Graves} G.~J., 2011, ArXiv e-prints: arXiv/1111.1070

\bibitem[{{Jee} {et~al}\mbox{.}(2013){Jee}, {Tyson}, {Schneider}, {Wittman},
  {Schmidt}, \& {Hilbert}}]{2013ApJ...765...74J}
{Jee} M.~J., {Tyson} J.~A., {Schneider} M.~D., {Wittman} D., {Schmidt} S.,
  {Hilbert} S., 2013, \apj, 765, 74

\bibitem[{{Kaiser}(1998)}]{1998ApJ...498...26K}
{Kaiser} N., 1998, \apj, 498, 26

\bibitem[{{Kaiser} \& {Squires}(1993)}]{1993ApJ...404..441K}
{Kaiser} N., {Squires} G., 1993, \apj, 404, 441

\bibitem[{{Lewis} \& {Challinor}(2006)}]{2006PhR...429....1L}
{Lewis} A., {Challinor} A., 2006, \physrep, 429, 1

\bibitem[{{Li} {et~al}\mbox{.}(2013){Li}, {Mo}, {Fan}, {Yang}, \&
  {Bosch}}]{2013MNRAS.430.3359L}
{Li} R., {Mo} H.~J., {Fan} Z., {Yang} X., {Bosch} F.~C.~v.~d., 2013, \mnras,
  430, 3359

\bibitem[{{Refregier}(2003)}]{2003ARA&A..41..645R}
{Refregier} A., 2003, \araa, 41, 645

\bibitem[{{Vallinotto} {et~al}\mbox{.}(2007){Vallinotto}, {Dodelson}, {Schimd},
  \& {Uzan}}]{2007PhRvD..75j3509V}
{Vallinotto} A., {Dodelson} S., {Schimd} C., {Uzan} J.-P., 2007, \prd, 75,
  103509

\end{thebibliography}

\end{document}